\documentclass[a4paper, amsfonts, amssymb, amsmath, reprint, showkeys, nofootinbib, twoside]{revtex4-1}
\usepackage[english]{babel}
\usepackage[utf8]{inputenc}
\usepackage[colorinlistoftodos, color=green!40, prependcaption]{todonotes}
\usepackage{amsthm}
\usepackage{ulem}
\usepackage{multirow} 
\usepackage{xcolor}
\usepackage{graphicx}
\usepackage{comment}
\usepackage{caption}
\usepackage[T1]{fontenc}

\usepackage[pdftex, pdftitle={Article}, pdfauthor={Author}]{hyperref} 
\definecolor{purple}{rgb}{0.58,0.0,0.83}

\definecolor{blue(pigment)}{rgb}{0.2, 0.2, 0.6}

\usepackage{scalerel}
\usepackage{tikz}
\usetikzlibrary{svg.path}

\definecolor{orcidlogocol}{HTML}{A6CE39}
\tikzset{
  orcidlogo/.pic={
    \fill[orcidlogocol] svg{M256,128c0,70.7-57.3,128-128,128C57.3,256,0,198.7,0,128C0,57.3,57.3,0,128,0C198.7,0,256,57.3,256,128z};
    \fill[white] svg{M86.3,186.2H70.9V79.1h15.4v48.4V186.2z}
                 svg{M108.9,79.1h41.6c39.6,0,57,28.3,57,53.6c0,27.5-21.5,53.6-56.8,53.6h-41.8V79.1z M124.3,172.4h24.5c34.9,0,42.9-26.5,42.9-39.7c0-21.5-13.7-39.7-43.7-39.7h-23.7V172.4z}
                 svg{M88.7,56.8c0,5.5-4.5,10.1-10.1,10.1c-5.6,0-10.1-4.6-10.1-10.1c0-5.6,4.5-10.1,10.1-10.1C84.2,46.7,88.7,51.3,88.7,56.8z};
  }
}

\newcommand\orcidicon[1]{\href{https://orcid.org/#1}{\mbox{\scalerel*{
\begin{tikzpicture}[yscale=-1,transform shape]
\pic{orcidlogo};
\end{tikzpicture}
}{|}}}}

\hypersetup{colorlinks=true,citecolor=blue,linkcolor=blue,urlcolor=blue}

\newcommand{\RA}{R_{A}}
\newcommand{\Pth}{P_{\text{th}}}
\newcommand{\Omk}{\Omega_{k}}

\begin{document}

\title{Does spatial curvature generate new thermodynamic criticality at the FLRW apparent horizon?}

\author{Samuel Lepe\orcidicon{0000-0002-3464-8337}}
\email{samuel.lepe@pucv.cl}

\author{Joel Saavedra\orcidicon{0000-0002-1430-3008}}
\email{joel.saavedra@pucv.cl}

\affiliation{Instituto de F\'\i sica, Pontificia Universidad Cat\'olica de Valpara\'\i so, Casilla 4950, Valpara\'\i so, Chile.}

\date{\today}

\begin{abstract}
We generalize the apparent-horizon thermodynamic construction of Ref.~\cite{Cruz:2023xjp} to a FLRW universe with non-zero spatial curvature, non-interacting cold dark matter and holographic-type dark energy. For nonzero curvature, the scale factor is an additional geometric variable in the horizon equation of state.
The usual criticality conditions are thus not well defined until a closure prescription for the curvature sector is provided.
We introduce a dimensionless curvature variable and restrict the thermodynamic variations to slices of constant value of this variable. For each such slice, a positive holographic coupling and nonlinear powers larger than one guaranty the existence of a unique positive critical point. The critical specific volume, temperature and pressure are shifted by the spatial curvature, while the critical ratio and the mean-field critical exponents remain unchanged. It thus rescales the critical quantities, without generating a new local universality class. We construct a Helmholtz potential out of the physical thermodynamic volume and the entropy conjugate to the rescaled horizon temperature. In the quadratic holographic model, the parametric coexistence curve which is equivalent to the Maxwell construction is derived from the equality of the Gibbs free energies of the competing branches. The associated latent heat disappears at the critical endpoint, resulting in a global first-order coexistence in the fixed-curvature ensemble. Whether a physical FLRW trajectory intersects the critical or coexistence locus is a separate dynamical question because the dimensionless curvature variable generally evolves during cosmological expansion.
\end{abstract}
\maketitle

\section{Introduction}

The relation between gravitational dynamics and horizon thermodynamics is especially clear in spherically symmetric spacetimes \cite{Bekenstein1973,Hawking:1975}. The Misner-Sharp energy \cite{MisnerSharp1964}, work density and energy-supply covector are placed into a geometric identity by the unified first law of Hayward \cite{Hayward1998}, which reduces to a first law when projected along a trapping horizon. For FLRW cosmology, the relevant causal surface is the apparent horizon, and the Friedmann equations can be written in a thermodynamic form if a temperature and entropy are assigned to the horizon \cite{CaiKim2005, Cai:2006rs,AkbarCai2007, Abdusattar2022}. This construction has been extended to equations of state,
critical phenomena, and modified gravitational scenarios
\cite{Kong2022, AbdusattarScalarTensor2023,Kong2023}. Related critical
phenomena have also been found in the thermodynamic description of
FLRW cosmologies with modified horizon entropy
\cite{Housset2024}.

This framework has motivated the study of phase behavior related to the cosmological apparent horizon. Gravitational systems can exhibit van der Waals-like criticality,
coexistence curves, and mean-field critical exponents
\cite{KubiznakMann2012, KubiznakMannTeo2017}.

Specifically, Ref. \cite{Cruz:2023xjp, Cruz:2025ebg} built a thermodynamic equation of state for a spatially flat FLRW universe with noninteracting cold dark matter and holographic-type dark energy and found van der Waals-like behavior for appropriate nonlinear energy densities. In that construction, spatial flatness is not a cosmetic assumption. For $k=0$, the apparent-horizon radius is $H\RA=1$, so that any dark-energy density expressed as a function of $H$ can be written directly in terms of the horizon radius. Conversely, for $k\ne0$, \begin{equation} H^2=\frac{1}{\RA^2}-\frac{k}{a^2}, \label{eq:HRA} \end{equation} and the scale factor is another geometric variable. Thus the pressure on the horizon is usually a function $\Pth=\Pth(T,v,a;k)$ and not only a function of $T$ and $v$.

The goal of this work is to clearly pose this problem and to extend the methodology of Ref.~\cite{Cruz:2023xjp} to a FLRW universe with non null spatial curvature. The relevant question is not simply whether one can repeat the flat-space computation for $k\ne0$, but whether spatial curvature generates a truly new thermodynamic interaction, or simply changes the characteristic scales of an existing critical point. To answer this question requires a precise prescription for closing the enlarged thermodynamic state space, otherwise the usual derivatives in relation to the specific volume are ambiguous.

To implement this closure we introduce the dimensionless curvature variable 
\begin{equation} 
\chi\equiv\frac{k\RA^2}{a^2}, 
\end{equation} 
and consider thermodynamic variations on slices of fixed $\chi$. In this ensemble the curvature dependence is absorbed in a $\chi$-dependent coefficient which multiplies the nonlinear holographic contribution. Therefore, the dimensional critical volume, temperature and pressure are affected by the spatial curvature, whereas the reduced equation of state and the critical exponents remain unchanged. This is a local statement about a given thermodynamic ensemble and should not be confused with evolution along a cosmological solution for which $\chi$ is generally time dependent.

A local inflection point in the equation of state does not by itself establish a global first-order phase transition. We form a Helmholtz potential in terms of the physical thermodynamic volume and the entropy conjugate to the rescaled horizon temperature. For the quadratic holographic model we find the coexistence curve from the equality of the Gibbs free energies and show that it is equivalent to the Maxwell construction in the pressure-volume plane. We also identify the spinodal branches and verify that the latent heat vanishes at the critical endpoint.

The rest of this paper is organized as follows. In
Sec.~\ref{sec:formalism}, we formulate apparent-horizon thermodynamics in nonflat FLRW spacetime, identify the geometric cancellation of the explicit curvature contributions, and introduce the fixed-curvature closure. In Sec.~\ref{sec:local-criticality}, we derive the holographic equation of state and analyze its critical point, curvature-dependent scales, reduced form, and critical exponents. In Sec.~\ref{sec:global-phase}, we construct the thermodynamic potentials and establish the global coexistence and spinodal structure. In Sec.~\ref{sec:cosmological-interpretation}, we examine the asymptotic branches and distinguish fixed-curvature thermodynamic variations from physical cosmological trajectories. We summarize our conclusions in
the final section. Throughout this work, we use units in which $8\pi G=c=1$. The horizon temperature is normalized in the same way as the pressure convention used below. The symbols $p_i$ denote the material pressures of the individual components, while $\Pth$ denotes the pressure-like thermodynamic variable obtained from the horizon work density.
\section{Horizon thermodynamics and state-space closure}
\label{sec:formalism}

\subsection{Apparent-horizon thermodynamics in nonflat FLRW spacetime}
\label{subsec:horizon-thermodynamics}

Consider the FLRW line element
\begin{equation}
 ds^2=-dt^2+a^2(t)
 \left[
 \frac{dr^2}{1-kr^2}+r^2d\Omega_2^2
 \right],
 \qquad
 k\in\{-1,0,+1\}.
 \label{eq:metric}
\end{equation}
Introducing the areal radius \(R=ar\), the metric can be written as
\begin{equation}
 ds^2=h_{ab}dx^a dx^b+R^2d\Omega_2^2,
\end{equation}
where \(x^a=(t,r)\). The apparent horizon is determined by
\begin{equation}
 h^{ab}\partial_aR\partial_bR=0,
\end{equation}
which gives
\begin{equation}
 \RA=\frac{1}{\sqrt{H^2+k/a^2}}.
 \label{eq:RA}
\end{equation}
The Einstein equations for a perfect fluid with total energy density
\(\rho\) and material pressure \(p\) are
\begin{align}
 H^2+\frac{k}{a^2}
 &=\frac{\rho}{3},
 \label{eq:F1}\\
 \dot H-\frac{k}{a^2}
 &=-\frac{1}{2}(\rho+p).
 \label{eq:F2}
\end{align}
It follows immediately that
\begin{equation}
 \rho=\frac{3}{\RA^2},
 \label{eq:rho-horizon}
\end{equation}
at the apparent horizon, independently of the sign of \(k\).

For a perfect fluid, the work density is defined by
\begin{equation}
 W
 \equiv
 -\frac{1}{2}T^{ab}h_{ab}
 =\frac{\rho-p}{2}.
 \label{eq:workdensity}
\end{equation}
The Misner-Sharp energy inside a sphere of areal radius \(R\) is
\begin{equation}
 E=
 \frac{R}{2G}
 \left(
 1-h^{ab}\partial_aR\partial_bR
 \right),
 \label{eq:misner-sharp}
\end{equation}
and Hayward's unified first law takes the form
\begin{equation}
 dE=A\Psi+W\,dV,
 \label{eq:UFL}
\end{equation}
where
\begin{equation}
 A=4\pi R^2,
 \qquad
 V=\frac{4\pi R^3}{3},
\end{equation}
and the energy-supply covector is
\begin{equation}
 \Psi_a=
 T_a{}^b\partial_bR
 +W\partial_aR.
 \label{eq:energy-supply}
\end{equation}
The covector \(\Psi_a\) describes the energy flux across surfaces of
constant areal radius. Projection of Eq.~\eqref{eq:UFL} along the
apparent horizon yields the corresponding horizon first law.

The dynamical surface gravity at the apparent horizon is
\begin{equation}
 \kappa_A=
 -\frac{1}{\RA}
 \left(
 1-\frac{\dot\RA}{2H\RA}
 \right).
 \label{eq:kappa}
\end{equation}
We restrict the analysis to the branch on which \(\kappa_A<0\) and
define the positive horizon temperature
\begin{equation}
 T_A=-\frac{\kappa_A}{2\pi}>0.
 \label{eq:horizon-temperature}
\end{equation}
This restriction avoids the piecewise thermodynamic description that
would be required if the prescription \(T_A=|\kappa_A|/(2\pi)\) were
used across a change in the sign of \(\kappa_A\).

To maintain the normalization adopted in Ref.~\cite{Cruz:2025ebg}, we use
the rescaled temperature
\begin{equation}
 T\equiv\pi T_A.
 \label{eq:Trescaled}
\end{equation}
This constant rescaling modifies numerical quantities involving the
temperature but does not affect the existence of the critical point or
its dependence on spatial curvature. We also introduce the horizon
specific volume
\begin{equation}
 v\equiv2\RA,
 \label{eq:specific-volume}
\end{equation}
so that the physical thermodynamic volume is
\begin{equation}
 V_A=\frac{4\pi\RA^3}{3}
 =\frac{\pi v^3}{6}.
 \label{eq:thermodynamic-volume}
\end{equation}

\subsection{Geometric cancellation of spatial curvature}
\label{subsec:curvature-cancellation}

Before specifying the matter model, it is useful to isolate the purely
geometric contribution of spatial curvature. Differentiating
Eq.~\eqref{eq:RA} gives
\begin{equation}
 \dot\RA
 =
 -H\RA^3
 \left(
 \dot H-\frac{k}{a^2}
 \right).
 \label{eq:RAdot}
\end{equation}
For a single perfect fluid, Eqs.~\eqref{eq:F1} and
\eqref{eq:F2} imply
\begin{equation}
 p=
 -2\left(
 \dot H-\frac{k}{a^2}
 \right)
 -\frac{3}{\RA^2}.
 \label{eq:pressure-single}
\end{equation}
The work density can therefore be written as
\begin{equation}
 W=
 \frac{3}{\RA^2}
 +\dot H-\frac{k}{a^2}
 =
 \frac{3}{\RA^2}
 -\frac{\dot\RA}{H\RA^3}.
 \label{eq:Wsingle}
\end{equation}
On the chosen temperature branch, Eqs.~\eqref{eq:kappa} and
\eqref{eq:horizon-temperature} give
\begin{equation}
 \frac{\dot\RA}{H\RA}
 =
 2\left(1-2\pi\RA T_A\right).
 \label{eq:RAdot-temperature}
\end{equation}
Substitution into Eq.~\eqref{eq:Wsingle} yields
\begin{equation}
 W=
 \frac{4\pi T_A}{\RA}
 +\frac{1}{\RA^2}
 =
 \frac{8T}{v}
 +\frac{4}{v^2}.
 \label{eq:singleEOS}
\end{equation}
All explicit dependence on \(k\) has cancelled. Consequently, spatial
curvature by itself does not generate the nonlinear interaction
required for a van der Waals-like inflection point.

We now consider two noninteracting components: cold dark matter and
dark energy,
\begin{equation}
 p_m=0,
 \qquad
 r\equiv\frac{\rho_m}{\rho_{\mathrm{de}}}.
 \label{eq:coincidence-parameter}
\end{equation}
The first Friedmann equation gives
\begin{equation}
 \frac{1}{1+r}
 =
 \frac{\rho_{\mathrm{de}}\RA^2}{3}.
 \label{eq:r-general}
\end{equation}
Following Ref.~\cite{Cruz:2023xjp}, the thermodynamic pressure associated
with the dark-energy sector is identified with its sector-resolved work
density,
\begin{equation}
 \Pth
 \equiv
 P_{\mathrm{de}}
 =
 \frac{\rho_{\mathrm{de}}-p_{\mathrm{de}}}{2}.
 \label{eq:Pdark}
\end{equation}
This quantity must be distinguished from the total work density,
\begin{equation}
 W=
 \frac{\rho_m+\rho_{\mathrm{de}}-p_{\mathrm{de}}}{2}
 =
 \Pth+\frac{\rho_m}{2}.
 \label{eq:W-Pdark}
\end{equation}
Using
\begin{equation}
 \rho_m=
 \frac{3}{\RA^2}-\rho_{\mathrm{de}},
\end{equation}
together with Eq.~\eqref{eq:singleEOS}, we obtain
\begin{align}
 \Pth
 &=
 W-\frac{\rho_m}{2}
 \nonumber\\
 &=
 \frac{4\pi T_A}{\RA}
 -\frac{1}{2\RA^2}
 +\frac{\rho_{\mathrm{de}}}{2}
 \nonumber\\
 &=
 \frac{4\pi T_A}{\RA}
 +\frac{1}{2\RA^2}
 \left(
 \frac{3}{1+r}-1
 \right).
 \label{eq:Ptwofluids}
\end{align}
Thus no explicit curvature term appears in the sector-resolved
equation of state either. Spatial curvature can modify the phase
structure only through the dependence of \(r\), or equivalently
\(\rho_{\mathrm{de}}\), on \(H\) and \(\RA\). This identifies the
nonlinear dark-energy sector, rather than the FLRW geometry alone, as
the origin of the nonideal contribution to the equation of state.

Spatial curvature may formally be represented as an effective fluid,
\begin{equation}
 \rho_k=-\frac{3k}{a^2},
 \qquad
 p_k=-\frac{1}{3}\rho_k.
 \label{eq:effective-curvature-fluid}
\end{equation}
This is only a bookkeeping convention. It does not convert curvature
into an independently conserved microscopic matter component and
should not be interpreted as a modification of the material equation
of state.

\subsection{Fixed-curvature thermodynamic closure}
\label{subsec:fixed-chi-closure}

Using \(v=2\RA\), Eq.~\eqref{eq:HRA} becomes
\begin{equation}
 H^2=
 \frac{4}{v^2}-\frac{k}{a^2}.
 \label{eq:Hva}
\end{equation}
A horizon relation derived from a dark-energy density
\(\rho_{\mathrm{de}}(H)\) therefore has the general form
\begin{equation}
 \Pth=\Pth(T,v,a;k),
 \label{eq:general-EOS}
\end{equation}
rather than \(\Pth=\Pth(T,v)\). Consequently, a derivative such as
\((\partial\Pth/\partial v)_T\) is incomplete unless the behavior of
the additional geometric variable is specified.

A convenient dimensionless curvature variable is
\begin{equation}
 \chi\equiv\frac{k\RA^2}{a^2}.
 \label{eq:chi}
\end{equation}
It satisfies
\begin{equation}
 H^2\RA^2=1-\chi,
 \qquad
 H^2=\frac{4(1-\chi)}{v^2}.
 \label{eq:Hchi}
\end{equation}
In terms of the conventional curvature density parameter
\begin{equation}
 \Omk\equiv-\frac{k}{a^2H^2},
 \label{eq:Omega-k}
\end{equation}
the relation between the two curvature variables is
\begin{equation}
 \chi=
 -\frac{\Omk}{1-\Omk},
 \qquad
 1-\chi=
 \frac{1}{1-\Omk}.
 \label{eq:chiOmega}
\end{equation}
The condition \(H^2>0\) requires
\begin{equation}
 1-\chi>0.
 \label{eq:physical-chi-domain}
\end{equation}
For a closed universe, \(k=+1\), one has \(0<\chi<1\). For an
open universe, \(k=-1\), one has \(\chi<0\), while \(\chi=0\)
corresponds to the spatially flat limit.

For completeness, the horizon temperature can also be expressed in
terms of the deceleration parameter
\begin{equation}
 q\equiv-1-\frac{\dot H}{H^2}.
\end{equation}
Using Eqs.~\eqref{eq:kappa}, \eqref{eq:RAdot}, and
\eqref{eq:Hchi}, one finds
\begin{equation}
 T_A=
 \frac{1}{4\pi\RA}
 \left[
 (1-q)(H\RA)^2+\chi
 \right].
 \label{eq:Tqchi}
\end{equation}
This expression must be evaluated on a dynamically consistent
cosmological solution. For example, a radiation-dominated universe
with nonzero spatial curvature satisfies
\begin{equation}
 q=
 1+\frac{k}{a^2H^2}
 =
 1-\Omk,
 \label{eq:qradiation}
\end{equation}
rather than \(q=1\). Assigning \(q=1\) and \(k\ne0\) independently
would therefore impose mutually inconsistent dynamical conditions and
could lead to a spurious conclusion about the sign of \(T_A\).

We close the thermodynamic state space by restricting the variations
to slices of fixed \(\chi\). The corresponding criticality conditions
are
\begin{equation}
 \left(
 \frac{\partial\Pth}{\partial v}
 \right)_{T,\chi}
 =0,
 \qquad
 \left(
 \frac{\partial^2\Pth}{\partial v^2}
 \right)_{T,\chi}
 =0.
 \label{eq:criticalconditions}
\end{equation}
These conditions define a well-posed two-dimensional thermodynamic
ensemble. They are not equivalent to varying \(v\) while keeping
\(a\) and \(k\) fixed. Indeed, for fixed nonzero \(k\),
\begin{equation}
 \chi=\frac{k v^2}{4a^2}=\text{constant}
 \qquad\Longrightarrow\qquad
 a\propto v.
 \label{eq:fixed-chi-scaling}
\end{equation}
The fixed-\(\chi\) prescription is therefore an ensemble closure and
not a cosmological evolution law. Along a physical FLRW trajectory,
both \(v\) and \(\chi\) generally evolve with time.

\section{Holographic equation of state and local criticality}
\label{sec:local-criticality}

\subsection{Holographic equation of state}
\label{subsec:holographic-EOS}

The holographic dark-energy framework originates from the
ultraviolet-infrared relation imposed on an effective quantum field
theory in the presence of gravity \cite{Cohen:1998zx}. Its cosmological
implementation relates the dark-energy density to an infrared cutoff,
whose standard choices include the Hubble scale and the future event
horizon \cite{Hsu:2004ri,Li:2004rb}. Motivated by this framework and
following Ref.~\cite{Cruz:2023xjp}, we consider the holographic-type
dark-energy density
\begin{equation}
 \rho_{\mathrm{de}}=3\alpha H^{2n},
 \label{eq:rhoDE}
\end{equation}
where
\begin{equation}
 [\alpha]=L^{2n-2}.
 \label{eq:alpha-dimension}
\end{equation}
For \(n=1\), the density has the standard Hubble-scale holographic
dependence proportional to \(H^2\). For \(n\ne1\), it should be
understood as a phenomenological holographic-type generalization,
rather than as the standard holographic dark-energy model.
Equation~\eqref{eq:r-general} then gives
\begin{equation}
 \frac{1}{1+r}
 =
 \alpha H^{2n}\RA^2.
 \label{eq:r-holographic-initial}
\end{equation}
Using
\begin{equation}
 H^2\RA^2=1-\chi,
\end{equation}
this relation becomes
\begin{equation}
 \frac{1}{1+r}
 =
 \alpha\RA^{-2(n-1)}(1-\chi)^n.
 \label{eq:r-holographic}
\end{equation}

Before imposing the fixed-\(\chi\) closure, substitution of
Eq.~\eqref{eq:rhoDE} into Eq.~\eqref{eq:Ptwofluids} yields
\begin{equation}
 \Pth(T,v,a;k)
 =
 \frac{8T}{v}
 -\frac{2}{v^2}
 +\frac{3\alpha}{2}
 \left(
 \frac{4}{v^2}-\frac{k}{a^2}
 \right)^n.
 \label{eq:EOSa}
\end{equation}
This expression makes explicit that, for \(k\ne0\), the horizon
pressure is not intrinsically a function of \(T\) and \(v\) alone.

On a fixed-\(\chi\) slice, Eq.~\eqref{eq:Hchi} reduces the equation of
state to
\begin{equation}
 \Pth(T,v;\chi)
 =
 \frac{8T}{v}
 -\frac{2}{v^2}
 +\frac{C_\chi}{v^{2n}},
 \label{eq:EOS}
\end{equation}
where
\begin{equation}
 C_\chi
 =
 6\alpha\,4^{n-1}(1-\chi)^n.
 \label{eq:Cchi}
\end{equation}
Spatial curvature therefore enters through a \(\chi\)-dependent
coefficient multiplying the nonlinear holographic term. It changes the
strength of this contribution on each fixed-\(\chi\) slice without
altering the functional dependence of the equation of state on \(T\)
and \(v\).

\subsection{Critical point and curvature shifts}
\label{subsec:critical-point}

The critical point is determined by
\begin{equation}
 \left(
 \frac{\partial\Pth}{\partial v}
 \right)_{T,\chi}
 =0,
 \qquad
 \left(
 \frac{\partial^2\Pth}{\partial v^2}
 \right)_{T,\chi}
 =0.
\end{equation}
Differentiating Eq.~\eqref{eq:EOS} at fixed \(T\) and \(\chi\) gives
\begin{align}
 \left(
 \frac{\partial\Pth}{\partial v}
 \right)_{T,\chi}
 &=
 -\frac{8T}{v^2}
 +\frac{4}{v^3}
 -\frac{2nC_\chi}{v^{2n+1}},
 \label{eq:dP}\\
 \left(
 \frac{\partial^2\Pth}{\partial v^2}
 \right)_{T,\chi}
 &=
 \frac{16T}{v^3}
 -\frac{12}{v^4}
 +\frac{2n(2n+1)C_\chi}{v^{2n+2}}.
 \label{eq:d2P}
\end{align}
Solving these equations simultaneously yields
\begin{equation}
 v_c^{2n-2}
 =
 3\alpha n(2n-1)4^{n-1}(1-\chi)^n.
 \label{eq:vcPower}
\end{equation}
The critical specific volume is therefore
\begin{equation}
 v_c
 =
 2
 \left[
 3\alpha n(2n-1)(1-\chi)^n
 \right]^{\frac{1}{2(n-1)}}.
 \label{eq:vc}
\end{equation}
The corresponding critical temperature and pressure are
\begin{align}
 T_c
 &=
 \frac{n-1}{2n-1}\frac{1}{v_c},
 \label{eq:Tc}\\
 P_c
 &=
 \frac{2(n-1)}{n}\frac{1}{v_c^2}.
 \label{eq:Pc}
\end{align}

For \(\alpha>0\), a real critical point with positive temperature,
pressure, and specific volume is obtained on the branch
\begin{equation}
 n>1,
 \qquad
 1-\chi>0.
 \label{eq:critical-domain}
\end{equation}
The second condition is already required by \(H^2>0\). Other
combinations of signs, particularly for noninteger \(n\), require a
separate analysis of reality, positivity, and thermodynamic stability
and will not be considered here.

The dimensionless critical ratio is
\begin{equation}
 \frac{P_cv_c}{T_c}
 =
 \frac{2(2n-1)}{n}.
 \label{eq:critical-ratio}
\end{equation}
It is independent of both \(\alpha\) and \(\chi\). Its numerical value
depends on the normalization adopted for \(T\) and \(\Pth\), but its
independence from spatial curvature does not.

For the quadratic model, \(n=2\), the critical quantities reduce to
\begin{equation}
 v_c=6\sqrt{2\alpha}\,(1-\chi),
 \qquad
 T_c=\frac{1}{3v_c},
 \qquad
 P_c=\frac{1}{v_c^2}.
 \label{eq:n2-critical}
\end{equation}
For \(\alpha=1\) and \(\chi=0\), one recovers
\begin{equation}
 v_c=6\sqrt{2},
\end{equation}
in agreement with the spatially flat result of
Ref.~\cite{Cruz:2025ebg}.

At fixed \(\alpha\) and \(n\), the displacement of the critical
specific volume relative to the flat case is
\begin{equation}
 \frac{v_c(\chi)}{v_c(0)}
 =
 (1-\chi)^{\frac{n}{2(n-1)}}.
 \label{eq:vcshift}
\end{equation}
For small curvature,
\begin{equation}
 \frac{\Delta v_c}{v_c}
 =
 -\frac{n}{2(n-1)}\chi
 +\mathcal O(\chi^2).
 \label{eq:vcshift-chi}
\end{equation}
Since
\begin{equation}
 \chi=
 -\frac{\Omk}{1-\Omk}
 =
 -\Omk+\mathcal O(\Omk^2),
\end{equation}
the same result can be written as
\begin{equation}
 \frac{\Delta v_c}{v_c}
 =
 \frac{n}{2(n-1)}\Omk
 +\mathcal O(\Omk^2).
 \label{eq:vcshift-Omega}
\end{equation}
Because \(T_c\propto v_c^{-1}\) and \(P_c\propto v_c^{-2}\),
one also obtains
\begin{align}
 \frac{\Delta T_c}{T_c}
 &=
 -\frac{n}{2(n-1)}\Omk
 +\mathcal O(\Omk^2),
 \label{eq:Tcshift}\\
 \frac{\Delta P_c}{P_c}
 &=
 -\frac{n}{n-1}\Omk
 +\mathcal O(\Omk^2).
 \label{eq:Pcshift}
\end{align}
For a closed universe, \(k=+1\) and \(\Omk<0\), so \(v_c\)
decreases while \(T_c\) and \(P_c\) increase relative to their flat
values. For an open universe, \(k=-1\) and \(\Omk>0\), so \(v_c\)
increases while \(T_c\) and \(P_c\) decrease. Thus, whenever
\(\lvert\Omk\rvert\ll1\), curvature produces only a small displacement
of the dimensional critical scales. The principal result is therefore
structural rather than an assertion of a large late-time
phenomenological effect.

\subsection{Reduced equation of state and critical exponents}
\label{subsec:reduced-criticality}

We introduce the reduced variables
\begin{equation}
 p\equiv\frac{\Pth}{P_c},
 \qquad
 \tau\equiv\frac{T}{T_c},
 \qquad
 \nu\equiv\frac{v}{v_c}.
 \label{eq:reduced}
\end{equation}
Using Eqs.~\eqref{eq:vcPower}-\eqref{eq:Pc}, the equation of state
becomes
\begin{equation}
 p(\tau,\nu)
 =
 \frac{4n}{2n-1}\frac{\tau}{\nu}
 -\frac{n}{n-1}\frac{1}{\nu^2}
 +\frac{1}{(n-1)(2n-1)}
 \frac{1}{\nu^{2n}}.
 \label{eq:reducedEOS}
\end{equation}
All dependence on \(\alpha\) and \(\chi\) has disappeared.
Consequently, spatial curvature changes the dimensional critical
scales but does not define a new reduced equation of state within the
fixed-\(\chi\) ensemble.

In the large-volume regime, \(\nu\gg1\), the reduced equation of state
approaches
\begin{equation}
 p\nu
 \simeq
 \frac{4n}{2n-1}\tau.
 \label{eq:ideal-limit}
\end{equation}
This relation has the functional dependence of an ideal gas, although
the coefficient differs from unity because the reduced variables are
normalized by the horizon critical point. In particular,
\begin{equation}
 \lim_{n\rightarrow\infty}p\nu=2\tau.
\end{equation}
The large-\(n\) limit therefore preserves the ideal-gas dependence but
not the conventionally normalized relation \(p\nu=\tau\).
The critical exponents characterize the leading singular behavior of
the thermodynamic response functions near the critical point. Defining
the reduced temperature
\begin{equation}
 t\equiv\frac{T-T_c}{T_c}=\tau-1,
 \label{eq:reduced-temperature-distance}
\end{equation}
we adopt the standard definitions
\begin{align}
 C_{V,\chi}
 &\sim |t|^{-\alpha_{\text{crit}}},
 \label{eq:alpha-definition}\\
 \eta
 &\sim (-t)^{\beta_{\text{crit}}},
 \qquad t<0,
 \label{eq:beta-definition}\\
 \kappa_{T,\chi}
 &\sim |t|^{-\gamma_{\text{crit}}},
 \label{eq:gamma-definition}\\
 |p-1|
 &\sim|\nu-1|^{\delta_{\text{crit}}},
 \qquad t=0.
 \label{eq:delta-definition}
\end{align}
Here, $\alpha_{\mathrm{crit}}$\footnote{The critical exponent
$\alpha_{\mathrm{crit}}$ is unrelated to the holographic coupling
$\alpha$ appearing in $\rho_{\mathrm{de}}=3\alpha H^{2n}$.}
denotes the critical exponent associated with the constant-volume
heat capacity. Now, 
\begin{equation}\label{eq:nu}
 \eta\equiv\nu_l-\nu_s,
\end{equation}
is the order parameter measuring the separation between the
coexisting large- and small-volume phases.
To determine the critical exponents, we first consider the entropy.
The Bekenstein-Hawking entropy of the apparent horizon is
\begin{equation}
 S_A=\frac{A}{4G}.
 \label{eq:horizon-entropy}
\end{equation}
In units \(8\pi G=1\), with \(A=\pi v^2\), this becomes
\begin{equation}
 S_A=2\pi^2v^2.
 \label{eq:entropy-v}
\end{equation}
Since the temperature used in the equation of state is \(T=\pi T_A\),
the entropy thermodynamically conjugate to \(T\) is
\begin{equation}
 \mathcal S\equiv\frac{S_A}{\pi}=2\pi v^2.
 \label{eq:rescaled-entropy}
\end{equation}
The heat capacity at fixed thermodynamic volume and fixed \(\chi\) is
therefore
\begin{equation}
 C_{V,\chi}
 \equiv
 T
 \left(
 \frac{\partial\mathcal S}{\partial T}
 \right)_{V_A,\chi}.
 \label{eq:CV-definition}
\end{equation}
Because
\begin{equation}
 V_A=\frac{\pi v^3}{6},
\end{equation}
holding \(V_A\) fixed also fixes \(v\). Consequently,
\begin{equation}
 C_{V,\chi}=0.
 \label{eq:CV-zero}
\end{equation}
There is no critical divergence in the constant-volume heat capacity,
and the corresponding exponent is
\begin{equation}
 \alpha_{\text{crit}}=0.
 \label{eq:alpha-critical}
\end{equation}

For the remaining exponents, we expand Eq.~\eqref{eq:reducedEOS} around
the critical point by writing
\begin{equation}
 \tau=1+\epsilon,
 \qquad
 \nu=1+\omega,
 \label{eq:critical-variables}
\end{equation}
where \(\lvert\epsilon\rvert\ll1\) and
\(\lvert\omega\rvert\ll1\). The reduced equation of state becomes
\begin{align}
 p={}&
 1
 +\frac{4n}{2n-1}\epsilon
 -\frac{4n}{2n-1}\epsilon\omega
 +\frac{4n}{2n-1}\epsilon\omega^2
 -\frac{2n}{3}\omega^3
 \nonumber\\
 &+
 \mathcal O
 \left(
 \omega^4,\epsilon\omega^3
 \right).
 \label{eq:critical-expansion}
\end{align}
The absence of terms proportional to \(\omega\) and \(\omega^2\) at
\(\epsilon=0\) follows from the criticality conditions.

Defining
\begin{equation}
 A_n\equiv\frac{4n}{2n-1},
 \qquad
 B_n\equiv\frac{2n}{3},
 \label{eq:An-Bn}
\end{equation}
the leading-order expansion reads
\begin{equation}
 p=
 1+A_n\epsilon
 -A_n\epsilon\omega
 -B_n\omega^3+\cdots.
 \label{eq:leading-critical-expansion}
\end{equation}
The near-critical coexistence branches are determined by equality of
the pressures,
\begin{equation}
 p(\epsilon,\omega_s)
 =
 p(\epsilon,\omega_l)
 =
 p_0,
 \label{eq:local-equal-pressure}
\end{equation}
together with the Maxwell equal-area condition in the physical
thermodynamic volume. Since
\begin{equation}
 \frac{V_A}{V_c}=\nu^3=(1+\omega)^3,
\end{equation}
the latter condition is
\begin{equation}
 \int_{\omega_s}^{\omega_l}
 \left[
 p(\epsilon,\omega)-p_0
 \right]
 (1+\omega)^2\,d\omega
 =0.
 \label{eq:local-maxwell}
\end{equation}
The factor \((1+\omega)^2\) follows from the physical volume measure
and can be replaced by unity only at leading order in the
near-critical expansion.
For \(\epsilon<0\), let \(\omega_s\) and \(\omega_l\) denote the
small- and large-volume coexistence branches. To leading order, the
equal-pressure and Maxwell conditions give
\begin{equation}
 \omega_l=-\omega_s\equiv\omega_0,
\end{equation}
with
\begin{equation}
 \omega_0^2
 =
 -\frac{A_n}{B_n}\epsilon
 =
 -\frac{6}{2n-1}\epsilon.
\end{equation}
The coexistence branches therefore satisfy
\begin{equation}
 \omega_l=
 \sqrt{-\frac{6\epsilon}{2n-1}},
 \qquad
 \omega_s=
 -\sqrt{-\frac{6\epsilon}{2n-1}}.
 \label{eq:coexistence-near-critical}
\end{equation}
Using the order parameter defined in
Eq.~\eqref{eq:nu}, one finds
\begin{equation}
 \eta
 =
 2\sqrt{-\frac{6\epsilon}{2n-1}}
 \propto(-\epsilon)^{1/2}.
\end{equation}
Hence,
\begin{equation}
 \beta_{\text{crit}}=\frac{1}{2}.
 \label{eq:beta-critical}
\end{equation}

The isothermal compressibility is
\begin{equation}
 \kappa_{T,\chi}
 \equiv
 -\frac{1}{V_A}
 \left(
 \frac{\partial V_A}{\partial\Pth}
 \right)_{T,\chi}.
 \label{eq:compressibility}
\end{equation}
Using \(V_A=V_c\nu^3\) and \(\Pth=P_cp\), it can be written as
\begin{equation}
 \kappa_{T,\chi}
 =
 -\frac{3}{P_c\nu}
 \left[
 \left(
 \frac{\partial p}{\partial\nu}
 \right)_\tau
 \right]^{-1}.
 \label{eq:reduced-compressibility}
\end{equation}
On the critical isochore, \(\omega=0\),
\begin{equation}
 \left(
 \frac{\partial p}{\partial\nu}
 \right)_\tau
 =
 -\frac{4n}{2n-1}\epsilon
 +\mathcal O(\epsilon^2).
\end{equation}
For \(\epsilon>0\), one therefore obtains
\begin{equation}
 \kappa_{T,\chi}
 \simeq
 \frac{3(2n-1)}{4nP_c}
 \frac{1}{\epsilon},
\end{equation}
which gives
\begin{equation}
 \gamma_{\text{crit}}=1.
 \label{eq:gamma-critical}
\end{equation}

On the critical isotherm, \(\epsilon=0\),
Eq.~\eqref{eq:critical-expansion} reduces to
\begin{equation}
 p-1
 =
 -\frac{2n}{3}\omega^3
 +\mathcal O(\omega^4).
\end{equation}
Since \(\omega=\nu-1\), it follows that
\begin{equation}
 |p-1|\propto|\nu-1|^3,
\end{equation}
and hence
\begin{equation}
 \delta_{\text{crit}}=3.
 \label{eq:delta-critical}
\end{equation}

The complete set of critical exponents is therefore
\begin{equation}
 \alpha_{\text{crit}}=0,
 \qquad
 \beta_{\text{crit}}=\frac{1}{2},
 \qquad
 \gamma_{\text{crit}}=1,
 \qquad
 \delta_{\text{crit}}=3.
 \label{eq:critical-exponents}
\end{equation}
These exponents are independent of \(\alpha\), \(\chi\), and \(n\),
although the corresponding critical amplitudes depend on \(n\). They
satisfy the standard mean-field scaling relations
\begin{align}
 \alpha_{\text{crit}}
 +2\beta_{\text{crit}}
 +\gamma_{\text{crit}}
 &=2,
 \\
 \gamma_{\text{crit}}
 &=
 \beta_{\text{crit}}
 \left(
 \delta_{\text{crit}}-1
 \right),
 \\
 \alpha_{\text{crit}}
 +\beta_{\text{crit}}
 \left(
 \delta_{\text{crit}}+1
 \right)
 &=2.
\end{align}
Thus, within the fixed-\(\chi\) ensemble, spatial curvature
renormalizes the dimensional critical scales but leaves the local
mean-field universality class unchanged.

\section{Thermodynamic potential and global phase structure}
\label{sec:global-phase}

The local inflection conditions establish the existence of a critical
point but do not, by themselves, demonstrate a global first-order phase
transition. To determine whether distinct thermodynamic branches can
coexist in equilibrium, we construct the Helmholtz and Gibbs
potentials on each fixed-\(\chi\) slice.

\subsection{Helmholtz and Gibbs free energies}
\label{subsec:thermodynamic-potentials}

The Helmholtz potential \(F(T,V_A;\chi)\) is defined by
\begin{equation}
 \Pth
 =
 -\left(
 \frac{\partial F}{\partial V_A}
 \right)_{T,\chi},
 \label{eq:helmholtz-definition}
\end{equation}
where
\begin{equation}
 V_A=\frac{\pi v^3}{6},
 \qquad
 dV_A=\frac{\pi v^2}{2}\,dv.
 \label{eq:volume-differential}
\end{equation}
Using the fixed-\(\chi\) equation of state,
\begin{equation}
 \Pth(T,v;\chi)
 =
 \frac{8T}{v}
 -\frac{2}{v^2}
 +\frac{C_\chi}{v^{2n}},
\end{equation}
Eq.~\eqref{eq:helmholtz-definition} gives
\begin{equation}
 \left(
 \frac{\partial F}{\partial v}
 \right)_{T,\chi}
 =
 -4\pi Tv
 +\pi
 -\frac{\pi C_\chi}{2}v^{2-2n}.
 \label{eq:F-derivative}
\end{equation}
For \(n\ne3/2\), integration yields
\begin{equation}
 F(T,v;\chi)
 =
 -2\pi Tv^2
 +\pi v
 +\frac{\pi C_\chi}{2(2n-3)}v^{3-2n}
 +F_0(\chi).
 \label{eq:helmholtz}
\end{equation}
The sign of the third term is positive for \(n>3/2\), while its
overall behavior for \(1<n<3/2\) is determined by both the denominator
and the positive power \(v^{3-2n}\).

The value \(n=3/2\) is regular in the equation of state and in the
critical quantities but is marginal in the integration of the
Helmholtz potential. In this case,
\begin{equation}
 F(T,v;\chi)
 =
 -2\pi Tv^2
 +\pi v
 -\frac{\pi C_\chi}{2}
 \ln\left(\frac{v}{v_0}\right)
 +F_0(\chi),
 \qquad
 n=\frac{3}{2},
 \label{eq:helmholtz-logarithmic}
\end{equation}
where \(v_0\) is an arbitrary reference scale. In the present analysis, \(n\) is treated as a real phenomenological
parameter on the expanding branch \(H>0\). The value \(n=3/2\) is
included only to make the integration of the thermodynamic potential
mathematically complete. If the holographic ansatz is restricted to
integer powers, as is customary in polynomial models, the physical
branch considered here begins at \(n=2\), and the logarithmic case is
absent.

More generally, direct integration of the pressure permits an
arbitrary function of \(T\) and \(\chi\). Thermodynamic consistency
with the entropy fixes its temperature dependence. Indeed,
\begin{equation}
 -
 \left(
 \frac{\partial F}{\partial T}
 \right)_{V_A,\chi}
 =
 2\pi v^2
 =
 \mathcal S,
 \label{eq:entropy-from-F}
\end{equation}
in agreement with Eq.~\eqref{eq:rescaled-entropy}. Since fixing \(V_A\)
also fixes \(v\), Eq.~\eqref{eq:entropy-from-F} requires the remaining
integration function to be independent of \(T\). The residual function
\(F_0(\chi)\) does not affect phase coexistence on a fixed-\(\chi\)
slice and will be set to zero in what follows.

The Gibbs free energy is
\begin{equation}
 G(T,\Pth;\chi)=F+\Pth V_A.
 \label{eq:gibbs-definition}
\end{equation}
For \(n\ne3/2\), its parametric representation in terms of \(v\) is
\begin{equation}
 G(T,v;\chi)
 =
 -\frac{2\pi}{3}Tv^2
 +\frac{2\pi}{3}v
 +\frac{\pi nC_\chi}{3(2n-3)}v^{3-2n}.
 \label{eq:gibbs-general}
\end{equation}
At fixed \(T\), \(\Pth\), and \(\chi\), different positive roots of
the equation of state represent competing thermodynamic branches.
A first-order transition occurs when the small- and large-volume
branches have the same Gibbs free energy.

\subsection{Maxwell construction and exact coexistence curve}
\label{subsec:coexistence}

Let \(v_s<v_l\) denote the specific volumes of the small- and
large-volume phases. Their thermodynamic volumes are
\begin{equation}
 V_s=\frac{\pi v_s^3}{6},
 \qquad
 V_l=\frac{\pi v_l^3}{6}.
 \label{eq:coexisting-volumes}
\end{equation}
Mechanical equilibrium requires
\begin{equation}
 \Pth(T,v_s;\chi)
 =
 \Pth(T,v_l;\chi)
 =
 P_0.
 \label{eq:equal-pressure}
\end{equation}
Thermodynamic coexistence additionally requires
\begin{equation}
 G(T,v_s;\chi)=G(T,v_l;\chi).
 \label{eq:equal-gibbs}
\end{equation}
Using \(dG=V_A\,d\Pth-\mathcal S\,dT\) at fixed \(\chi\),
Eq.~\eqref{eq:equal-gibbs} is equivalent to the Maxwell
equal-area condition
\begin{equation}
 \int_{V_s}^{V_l}
 \Pth(T,V_A;\chi)\,dV_A
 =
 P_0(V_l-V_s).
 \label{eq:maxwell-volume}
\end{equation}
In terms of \(v\), this becomes
\begin{equation}
 \frac{\pi}{2}
 \int_{v_s}^{v_l}
 \Pth(T,v;\chi)v^2\,dv
 =
 \frac{\pi P_0}{6}
 \left(v_l^3-v_s^3\right).
 \label{eq:maxwell-v}
\end{equation}
The factor \(v^2\) in the integration measure is essential. The
condition
\begin{equation}
 \int_{v_s}^{v_l}\Pth\,dv
 =
 P_0(v_l-v_s)
\end{equation}
would incorrectly treat the specific volume \(v\) as the extensive
thermodynamic volume and is not equivalent to
Eq.~\eqref{eq:maxwell-volume}.

For the quadratic holographic model, \(n=2\), the reduced equation of
state is
\begin{equation}
 p(\tau,\nu)
 =
 \frac{8\tau}{3\nu}
 -\frac{2}{\nu^2}
 +\frac{1}{3\nu^4}.
 \label{eq:n2-reduced-EOS}
\end{equation}
Introduce the parameter
\begin{equation}
 x\equiv\frac{\nu_s}{\nu_l},
 \qquad
 0<x\leq1.
 \label{eq:coexistence-parameter}
\end{equation}
Solving the equal-pressure and Maxwell conditions gives the exact
parametric coexistence curve
\begin{align}
 \nu_l(x)
 &=
 \frac{\sqrt{x^2+4x+1}}{\sqrt{6}\,x},
 \label{eq:nul-parametric}\\
 \nu_s(x)
 &=
 \frac{\sqrt{x^2+4x+1}}{\sqrt{6}},
 \label{eq:nus-parametric}\\
 \tau(x)
 &=
 \frac{3\sqrt{6}\,x(x+1)}
 {(x^2+4x+1)^{3/2}},
 \label{eq:tau-parametric}\\
 p_0(x)
 &=
 \frac{36x^2}{(x^2+4x+1)^2}.
 \label{eq:p-parametric}
\end{align}
These reduced coexistence relations are independent of \(\alpha\) and
\(\chi\), consistently with the curvature independence of the reduced
equation of state.

In the critical limit,
\begin{equation}
 x\longrightarrow1:
 \qquad
 \nu_s,\nu_l,\tau,p_0\longrightarrow1.
 \label{eq:critical-coexistence-limit}
\end{equation}
At the low-temperature end,
\begin{equation}
 x\longrightarrow0:
 \qquad
 \tau\longrightarrow0,
 \quad
 p_0\longrightarrow0,
 \quad
 \nu_l\longrightarrow\infty,
 \quad
 \nu_s\longrightarrow\frac{1}{\sqrt6}.
 \label{eq:low-temperature-limit}
\end{equation}

The entropy discontinuity between the two phases is
\begin{equation}
 \Delta\mathcal S
 =
 \mathcal S_l-\mathcal S_s
 =
 2\pi\left(v_l^2-v_s^2\right)>0.
 \label{eq:entropy-discontinuity}
\end{equation}
The latent heat is consequently
\begin{equation}
 Q_{\mathrm{lat}}
 =
 T\Delta\mathcal S.
 \label{eq:latent-heat}
\end{equation}
Both \(\Delta\mathcal S\) and \(Q_{\mathrm{lat}}\) vanish continuously
as \(x\to1\), showing that the first-order coexistence curve terminates
at the local critical point.

\begin{figure*}[t]
 \centering
 \includegraphics[width=0.96\textwidth]
 {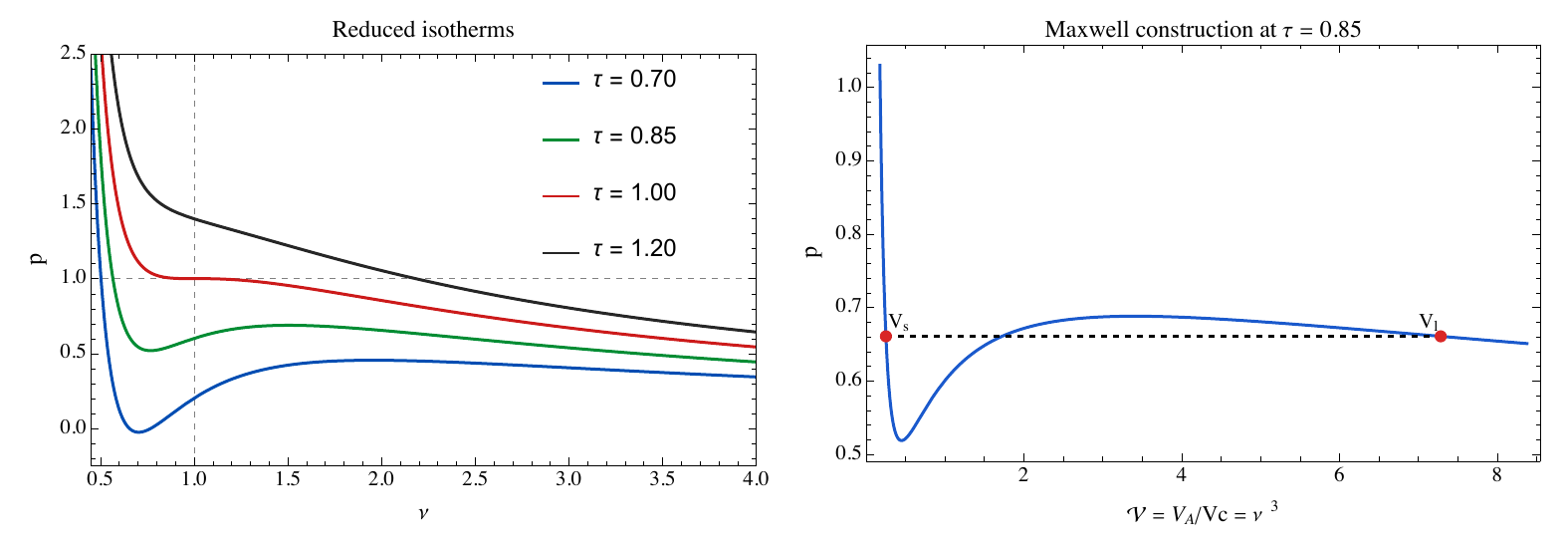}
 \caption{
 Reduced phase structure for the quadratic holographic model,
 \(n=2\). The left panel shows the reduced isotherms below, at,
 and above the critical temperature. For \(\tau<1\), two turning
 points delimit the mechanically unstable branch. At \(\tau=1\),
 they merge at the critical point \((\nu,p)=(1,1)\). The right
 panel shows the Maxwell construction at \(\tau=0.85\) in terms
 of the reduced thermodynamic volume
 \(\mathcal V=V_A/V_c=\nu^3\). The horizontal line \(p=p_0\)
 connects the coexisting phases and satisfies the equal-area
 condition in the \(p\)-\(\mathcal V\) plane.
 }
 \label{fig:isotherms-maxwell}
\end{figure*}

\begin{figure*}[!t]
 \centering
 \includegraphics[width=0.47\textwidth]
 {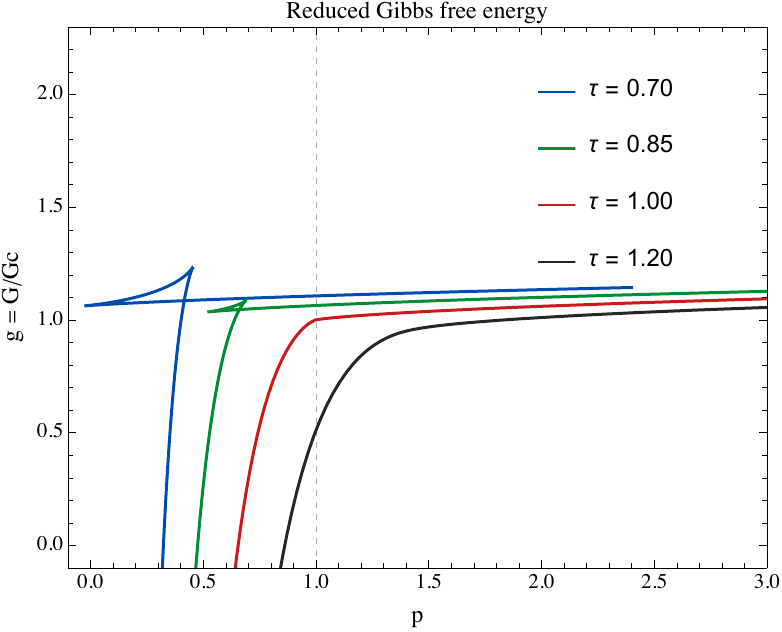}
 \hfill
 \includegraphics[width=0.51\textwidth]
 {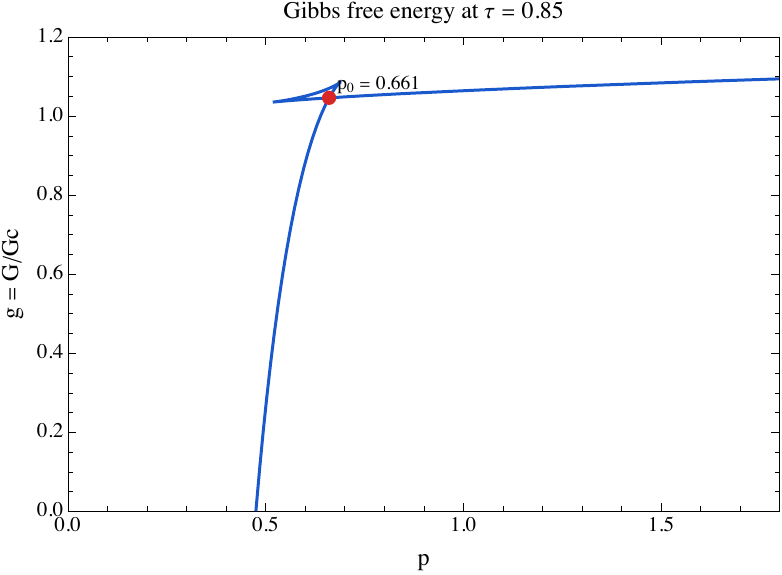}
 \caption{
 Gibbs free-energy structure for the quadratic holographic
 model, \(n=2\). The left panel shows the reduced Gibbs free
 energy parametrized by the reduced pressure for representative
 temperatures. For \(\tau<1\), the multibranch structure permits
 distinct phases at the same temperature and pressure. The right
 panel shows the coexistence condition at \(\tau=0.85\). The
 competing branches have equal Gibbs free energy at
 \(p_0\simeq0.661\), consistently with the Maxwell construction
 performed using the physical thermodynamic volume
 \(V_A=\pi v^3/6\).
 }
 \label{fig:gibbs-structure}
\end{figure*}

For \(n=2\), setting \(F_0(\chi)=0\) and evaluating
Eq.~\eqref{eq:gibbs-general} at the critical point gives
\begin{equation}
 G_c=\frac{2\pi v_c}{3}.
 \label{eq:critical-gibbs}
\end{equation}
The reduced Gibbs free energy is therefore
\begin{equation}
 g(\tau,\nu)
 \equiv
 \frac{G}{G_c}
 =
 \nu-\frac{\tau\nu^2}{3}
 +\frac{1}{3\nu}.
 \label{eq:reduced-gibbs}
\end{equation}
At fixed \(\tau\), the Gibbs curve is represented parametrically by
\begin{equation}
 \left\{
 p(\tau,\nu),g(\tau,\nu)
 \right\}.
 \label{eq:parametric-gibbs}
\end{equation}
The coexistence condition is
\begin{equation}
 p(\tau,\nu_s)
 =
 p(\tau,\nu_l)
 =
 p_0,
 \qquad
 g(\tau,\nu_s)
 =
 g(\tau,\nu_l).
 \label{eq:reduced-coexistence-condition}
\end{equation}
For \(\tau=0.85\), the Maxwell construction shown in the right panel
of Fig.~\ref{fig:isotherms-maxwell} gives \(p_0\simeq0.661\).
Independently, the Gibbs branches displayed in the right panel of
Fig.~\ref{fig:gibbs-structure} intersect at the same pressure. The
agreement between these two constructions verifies the global
coexistence condition. The corresponding global coexistence construction for the spatially
flat model was developed in Ref.~\cite{Cruz:2025ebg}.

\subsection{Spinodal curves and local stability}
\label{subsec:spinodal}

The limits of local mechanical stability are determined by the
spinodal condition
\begin{equation}
 \left(
 \frac{\partial p}{\partial\nu}
 \right)_\tau
 =0.
 \label{eq:spinodal-condition}
\end{equation}
For the general reduced equation of state,
\begin{equation}
 \left(
 \frac{\partial p}{\partial\nu}
 \right)_\tau
 =
 -\frac{4n}{2n-1}\frac{\tau}{\nu^2}
 +\frac{2n}{n-1}\frac{1}{\nu^3}
 -\frac{2n}{(n-1)(2n-1)}
 \frac{1}{\nu^{2n+1}}.
 \label{eq:spinodal-derivative}
\end{equation}
Solving Eq.~\eqref{eq:spinodal-condition} for the temperature gives
\begin{equation}
 \tau_{\mathrm{sp}}(\nu)
 =
 \frac{(2n-1)\nu^{2n-2}-1}
 {2(n-1)\nu^{2n-1}}.
 \label{eq:spinodal-temperature}
\end{equation}
Substitution into Eq.~\eqref{eq:reducedEOS} yields
\begin{equation}
 p_{\mathrm{sp}}(\nu)
 =
 \frac{n\nu^{2n-2}-1}
 {(n-1)\nu^{2n}}.
 \label{eq:spinodal-pressure}
\end{equation}
Both branches meet at
\begin{equation}
 \nu=1,
 \qquad
 \tau_{\mathrm{sp}}=1,
 \qquad
 p_{\mathrm{sp}}=1.
 \label{eq:spinodal-critical-point}
\end{equation}

Positive spinodal temperature and pressure require, respectively,
\begin{equation}
 \nu>
 \left(\frac{1}{2n-1}\right)^{\frac{1}{2n-2}},
 \qquad
 \nu>
 n^{-\frac{1}{2n-2}}.
 \label{eq:positive-spinodal-domain}
\end{equation}
For \(n=2\), the spinodal curve reduces to
\begin{equation}
 \tau_{\mathrm{sp}}(\nu)
 =
 \frac{3\nu^2-1}{2\nu^3},
 \qquad
 p_{\mathrm{sp}}(\nu)
 =
 \frac{2\nu^2-1}{\nu^4}.
 \label{eq:n2-spinodal}
\end{equation}
In the positive-pressure region, the small-volume spinodal branch
satisfies
\begin{equation}
 \frac{1}{\sqrt2}<\nu<1,
\end{equation}
while the large-volume branch satisfies
\begin{equation}
 \nu>1.
\end{equation}

The isothermal compressibility can be written as
\begin{equation}
 \kappa_{T,\chi}
 =
 -\frac{3}{P_c\nu}
 \left[
 \left(
 \frac{\partial p}{\partial\nu}
 \right)_\tau
 \right]^{-1}.
\end{equation}
Consequently, local mechanical stability requires
\begin{equation}
 \left(
 \frac{\partial p}{\partial\nu}
 \right)_\tau<0.
 \label{eq:mechanical-stability}
\end{equation}
A branch with positive slope is mechanically unstable, and
\(\kappa_{T,\chi}\) diverges at the spinodal boundary.

The heat capacity at fixed pressure and fixed \(\chi\) is
\begin{equation}
 C_{P,\chi}
 \equiv
 T
 \left(
 \frac{\partial\mathcal S}{\partial T}
 \right)_{\Pth,\chi}.
 \label{eq:CP-definition}
\end{equation}
Using \(\mathcal S=2\pi v^2\) and the equation of state, one finds
\begin{equation}
 C_{P,\chi}
 =
 -\frac{32\pi T}
 {\left(\partial\Pth/\partial v\right)_{T,\chi}}.
 \label{eq:CP}
\end{equation}
Thus \(C_{P,\chi}\) is positive on mechanically stable branches,
negative on the unstable branch, and diverges on the spinodal curve.

\begin{figure}[!t]
 \centering
 \includegraphics[width=\columnwidth]
 {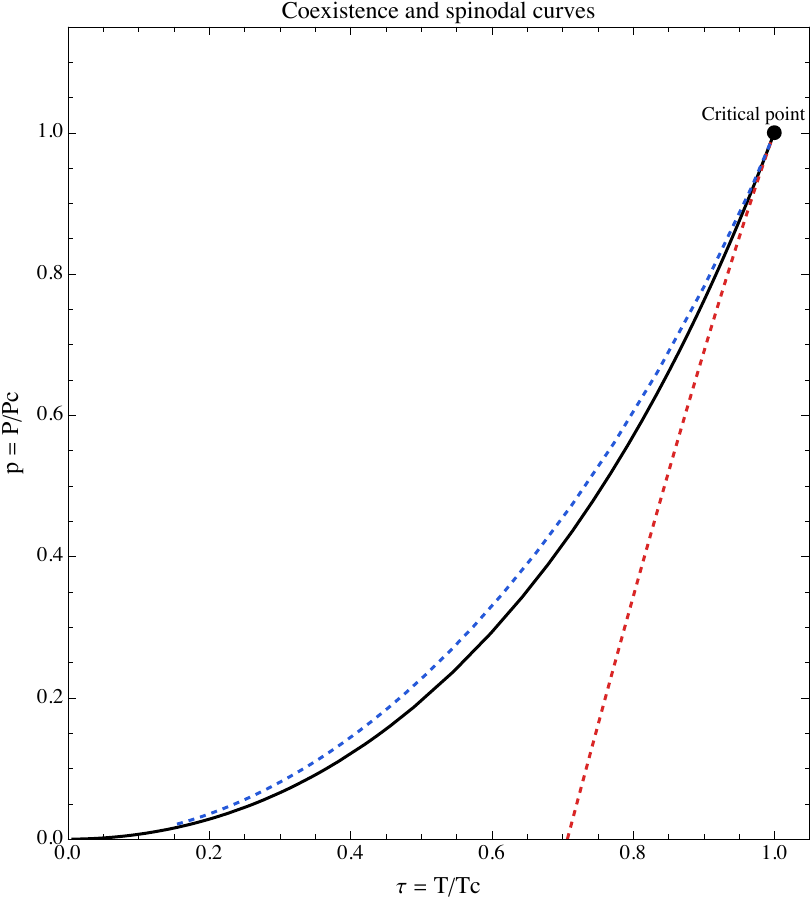}
 \caption{
 Reduced phase diagram in the \((\tau,p)\) plane for the
 quadratic holographic model, \(n=2\). The solid black curve is
 the first-order coexistence line obtained from equality of the
 Gibbs free energies, equivalently from the Maxwell construction
 in the \(\Pth\)-\(V_A\) plane. The red dashed curve is the
 small-volume spinodal branch, while the blue dashed curve is the
 large-volume spinodal branch. All three curves terminate at the
 critical point \((\tau,p)=(1,1)\).
 }
 \label{fig:coexistence-spinodal}
\end{figure}

The coexistence curve lies between the two spinodal branches. The
region bounded by the spinodal curves is mechanically unstable, while
the regions between the coexistence curve and each spinodal branch
represent metastable continuations of the small- and large-volume
phases. The reduced coexistence and spinodal curves are independent of
\(\alpha\) and \(\chi\). Spatial curvature changes their dimensional
scales through \(v_c\), \(T_c\), and \(P_c\), but does not modify the
global reduced phase structure.
\section{Cosmological interpretation}
\label{sec:cosmological-interpretation}

The fixed-\(\chi\) construction defines a consistent thermodynamic
ensemble, but it does not by itself describe the time evolution of an
FLRW universe. In this section, we examine the relation between the
thermodynamic state space and physical cosmological trajectories,
focusing first on the asymptotic branches of the quadratic
holographic model.

\subsection{Asymptotic branches of the quadratic model}
\label{subsec:asymptotic-branches}

The limit \(a\to\infty\) requires special care because the curvature
variable depends on the combination \(aH\), rather than on the scale
factor alone. From
\begin{equation}
 \chi=\frac{k\RA^2}{a^2}
\end{equation}
and
\begin{equation}
 \RA^{-2}=H^2+\frac{k}{a^2},
\end{equation}
one obtains
\begin{equation}
 \chi=\frac{k}{(aH)^2+k}.
 \label{eq:chi-aH}
\end{equation}
Consequently, the asymptotic behavior of \(\chi\) is controlled by
\(aH\).

To illustrate the possible branches, consider \(n=2\), noninteracting
cold dark matter, and
\begin{equation}
 \rho_m(a)
 =
 \rho_{m0}
 \left(
 \frac{a_0}{a}
 \right)^3.
 \label{eq:matter-density}
\end{equation}
The first Friedmann equation becomes
\begin{equation}
 \alpha H^4-H^2+f(a)=0,
 \label{eq:quadratic-H}
\end{equation}
where
\begin{equation}
 f(a)
 =
 \frac{\rho_{m0}}{3}
 \left(
 \frac{a_0}{a}
 \right)^3
 -\frac{k}{a^2}.
 \label{eq:f-a}
\end{equation}
The two algebraic branches are
\begin{equation}
 H_\pm^2
 =
 \frac{1\pm\sqrt{1-4\alpha f(a)}}{2\alpha}.
 \label{eq:H-branches}
\end{equation}
Their physical admissibility requires
\begin{equation}
 1-4\alpha f(a)\geq0,
 \qquad
 H_\pm^2\geq0.
 \label{eq:branch-reality}
\end{equation}

For \(\alpha>0\), the plus branch approaches a de Sitter regime:
\begin{equation}
 H_+^2\longrightarrow\frac{1}{\alpha},
 \qquad
 aH_+\longrightarrow\infty,
 \qquad
 \chi\longrightarrow0.
 \label{eq:plus-asymptotic}
\end{equation}
Thus the relative contribution of spatial curvature vanishes on this
branch, and
\begin{equation}
 C_\chi\longrightarrow24\alpha
 \qquad
 (n=2).
 \label{eq:Cchi-plus}
\end{equation}

For small \(f(a)\), the minus branch behaves as
\begin{equation}
 H_-^2=f(a)+\mathcal O\bigl(f(a)^2\bigr).
 \label{eq:minus-small-f}
\end{equation}
Its asymptotic behavior depends on the sign of \(k\).

For a closed universe, \(k=+1\),
\begin{equation}
 f(a)
 =
 \frac{\rho_{m0}a_0^3}{3a^3}
 -\frac{1}{a^2}
 \longrightarrow
 -\frac{1}{a^2}.
\end{equation}
At sufficiently large \(a\), one has \(f(a)<0\), and therefore
\(H_-^2<0\). The minus branch is consequently not physically
admissible in the asymptotic closed-universe regime.

For an open universe, \(k=-1\),
\begin{equation}
 f(a)
 =
 \frac{1}{a^2}
 +\frac{\rho_{m0}a_0^3}{3a^3},
 \label{eq:f-open}
\end{equation}
so that
\begin{equation}
 H_-^2
 =
 \frac{1}{a^2}
 +\frac{\rho_{m0}a_0^3}{3a^3}
 +\mathcal O(a^{-4}).
 \label{eq:Hminus-open}
\end{equation}
It follows that
\begin{equation}
 aH_-\longrightarrow1^+,
 \label{eq:Milne-limit}
\end{equation}
which corresponds to a Milne-like asymptotic regime. Using
Eq.~\eqref{eq:chi-aH}, one obtains
\begin{equation}
 \chi
 =
 -\frac{1}{(aH_-)^2-1}
 \longrightarrow-\infty.
 \label{eq:chi-Milne}
\end{equation}
Thus the open Milne-like branch does not approach \(\chi=0\), even
though \(k/a^2\to0\) in absolute magnitude.

For \(n=2\), the coefficient in the fixed-\(\chi\) equation of state
behaves as
\begin{equation}
 C_\chi=24\alpha(1-\chi)^2,
\end{equation}
and therefore diverges when \(\chi\to-\infty\). This divergence does
not imply that the physical holographic contribution diverges. Indeed,
\begin{equation}
 \frac{C_\chi}{v^4}
 =
 \frac{3\alpha}{2}H^4
 \longrightarrow0
 \label{eq:physical-Milne}
\end{equation}
on the Milne-like branch. The coefficient \(C_\chi\) cannot therefore
be interpreted independently of the accompanying factor \(v^{-4}\).

This example also shows that \(a\to\infty\) does not, by itself,
guarantee that curvature is dynamically negligible. The relevant
quantity is
\begin{equation}
 \frac{|k|}{a^2H^2}=|\Omk|.
 \label{eq:relative-curvature}
\end{equation}
Curvature becomes negligible relative to a positive cosmological
constant or an asymptotically de Sitter component, but it can remain
important relative to pressureless matter in an open universe without
a late-time accelerating contribution.

The Milne-like limit discussed here refers only to the open,
curvature-dominated FLRW geometry, for which \(a(t)\propto t\) and
\(q\to0\). It should not be identified with the specific Dirac-Milne
matter-antimatter cosmology \cite{Benoit-Levy:2011esx}, which involves additional assumptions
about the gravitational behavior of matter and antimatter.

\subsection{Thermodynamic slices and cosmological trajectories}
\label{subsec:thermodynamic-trajectories}

The fixed-\(\chi\) criticality analysis considers independent
thermodynamic variations of \(T\) and \(v\) while holding \(\chi\)
constant. Along a physical FLRW solution, however,
\begin{equation}
 T=T(t),
 \qquad
 v=v(t),
 \qquad
 \chi=\chi(t).
 \label{eq:trajectory-variables}
\end{equation}
A cosmological history therefore defines a one-dimensional curve in
the enlarged state space \((T,v,\chi)\), whereas the thermodynamic
equation of state defines a higher-dimensional equilibrium surface.

For a curve \(\chi=\chi(v)\) contained in an isothermal section, the
pressure derivative is
\begin{equation}
 \frac{d\Pth}{dv}\bigg|_{T,\mathrm{path}}
 =
 \left(
 \frac{\partial\Pth}{\partial v}
 \right)_{T,\chi}
 +
 \left(
 \frac{\partial\Pth}{\partial\chi}
 \right)_{T,v}
 \frac{d\chi}{dv}.
 \label{eq:total-derivative}
\end{equation}
The corresponding second derivative is
\begin{align}
 \frac{d^2\Pth}{dv^2}\bigg|_{T,\mathrm{path}}
 ={}&
\partial_{vv}P_{th}
 +2 \partial_{v\chi}P_{th}\chi'
 +\partial_{\chi \chi}P_{th}(\chi')^2
 +\partial_{\chi} P_{th}\chi'',
 \label{eq:second-total-derivative}
\end{align}
where primes denote derivatives with respect to $(v)$ along the
selected path and all partial derivatives are evaluated at fixed
values of the remaining state variables. Therefore, a critical point
defined by
\begin{equation}
 \partial_vP_{th}=0,
 \qquad
 \partial_{vv}P_{th}=0,
\end{equation}
at fixed \(\chi\) is not automatically a critical point of a path
along which \(\chi\) varies.

A physical cosmological trajectory is not generally isothermal. Its
pressure evolution is instead
\begin{equation}
 \dot\Pth
 =
 \left(
 \frac{\partial\Pth}{\partial T}
 \right)_{v,\chi}\dot T
 +
 \left(
 \frac{\partial\Pth}{\partial v}
 \right)_{T,\chi}\dot v
 +
 \left(
 \frac{\partial\Pth}{\partial\chi}
 \right)_{T,v}\dot\chi.
 \label{eq:pressure-time-evolution}
\end{equation}
For the fixed-\(\chi\) equation of state,
\begin{equation}
 \left(
 \frac{\partial\Pth}{\partial\chi}
 \right)_{T,v}
 =
 -\frac{nC_\chi}{(1-\chi)v^{2n}}.
 \label{eq:P-chi-derivative}
\end{equation}
The last term in Eq.~\eqref{eq:pressure-time-evolution} explicitly
measures the departure of the cosmological evolution from a
fixed-\(\chi\) thermodynamic slice.

For \(k\ne0\), differentiation of
\(\chi=k\RA^2/a^2\) gives
\begin{equation}
 \frac{\dot\chi}{\chi}
 =
 2\frac{\dot\RA}{\RA}-2H.
 \label{eq:chi-dot}
\end{equation}
Since \(v=2\RA\), this can equivalently be written as
\begin{equation}
 \frac{\dot\chi}{\chi}
 =
 2\frac{\dot v}{v}-2H.
 \label{eq:chi-dot-v}
\end{equation}
A constant-\(\chi\) evolution would require
\begin{equation}
 \frac{\dot\RA}{\RA}=H,
 \qquad\text{or equivalently}\qquad
 \RA\propto a,
 \label{eq:constant-chi-dynamics}
\end{equation}
which is a special dynamical condition and is not satisfied by a
generic FLRW solution.

The thermodynamic and cosmological questions are therefore distinct:
\begin{enumerate}
 \item Does the fixed-\(\chi\) equilibrium state space contain a
 critical point and a first-order coexistence curve?

 \item Does a dynamically admissible FLRW solution intersect the
 corresponding critical or coexistence locus?
\end{enumerate}
The first question is answered by the local and global thermodynamic
analysis developed in Secs.~\ref{sec:local-criticality} and
\ref{sec:global-phase}. The second requires solving the Friedmann and
continuity equations for the selected matter content and initial
conditions.

Even an intersection between a cosmological trajectory and the
coexistence surface is only a necessary condition for interpreting the
transition as a physical event. The equilibrium construction also
assumes that the relevant relaxation time is sufficiently short
compared with the cosmological expansion timescale. Establishing this
condition would require a dynamical description of perturbations,
nucleation, or relaxation processes beyond the equilibrium equation
of state.

Finally, a density containing additional kinematic variables, for
example
\begin{equation}
 \rho_{\mathrm{de}}
 =
 3\left(
 \alpha H^{2n}+\beta\dot H
 \right),
 \label{eq:extended-density}
\end{equation}
would enlarge the state space once again. In such a model, \(\dot H\)
must be eliminated through a specified dynamical closure, and
derivatives of \(\chi\) may enter explicitly. This extension cannot be
implemented consistently by direct substitution into
Eq.~\eqref{eq:EOS} and should be analyzed as a separate thermodynamic
system.

\section{Conclusions}
\label{sec:conclusions}
We have investigated the effect of spatial curvature on
apparent-horizon thermodynamic criticality in Einstein-FLRW cosmology
with noninteracting cold dark matter and holographic-type dark energy.
For nonzero curvature, the scale factor remains an independent
geometric variable in the horizon equation of state. Consequently,
critical derivatives with respect to the specific volume are not
uniquely defined until a closure prescription for the curvature sector
is specified.

We showed that the explicit curvature contributions cancel from the
work-density equation of state before the holographic model is
introduced. Spatial curvature alone therefore does not generate the
nonlinear interaction responsible for the critical behavior. Curvature
reenters indirectly through the dependence of the holographic density
on the Hubble parameter.

We closed the enlarged thermodynamic state space by introducing a
dimensionless curvature variable and restricting the variations to
slices on which this variable is held fixed. Within this ensemble,
spatial curvature modifies the dimensional critical volume,
temperature, and pressure but leaves the reduced equation of state
unchanged. The dimensionless critical ratio and the mean-field critical
exponents are consequently independent of curvature. Spatial curvature
therefore renormalizes the critical scales without producing a new
local universality class.

Using the physical thermodynamic volume, we constructed a Helmholtz
potential consistent with the Bekenstein-Hawking entropy and obtained
the corresponding Gibbs free energy. For the quadratic holographic
model, equality of the Gibbs free energies is equivalent to the Maxwell
construction in the pressure-volume plane and determines an exact
parametric first-order coexistence curve. The coexistence line lies
between the two spinodal branches, and all three curves terminate at
the critical point. The latent heat vanishes continuously at this
endpoint. These results establish a global equilibrium phase structure
within the fixed-curvature ensemble.

The thermodynamic phase diagram must nevertheless be distinguished
from the evolution of a physical FLRW universe. Along a cosmological
solution, the temperature, apparent-horizon radius, and dimensionless
curvature variable generally evolve simultaneously. A fixed-curvature
critical point is therefore not automatically reached by a
cosmological trajectory. Determining whether the universe intersects
the critical or coexistence locus requires solving the complete
dynamical system for specified matter content and initial conditions.
Even such an intersection would provide only a necessary condition for
a physical transition, because the relevant equilibration and
cosmological expansion timescales must also be compared.

\begin{acknowledgments}
S.L. acknowledges support from FONDECYT Grant No.~1250969, Chile. J. Saavedra acknowledges the financial support of Fondecyt Grant 1220065.
\end{acknowledgments}

\appendix

\section{Algebraic derivation of the critical quantities}
\label{app:critical_quantities}

The fixed-$\chi$ equation of state is
\begin{equation}
    P(T,v;\chi)
    =
    \frac{8T}{v}
    -
    \frac{2}{v^2}
    +
    \frac{C_\chi}{v^{2n}}.
    \label{eq:appendix_eos}
\end{equation}
At the critical point, define the dimensionless quantity
\begin{equation}
    X
    \equiv
    C_\chi v_c^{2-2n}.
    \label{eq:X_definition}
\end{equation}
Multiplying the first criticality condition by $v_c^3$ gives
\begin{equation}
    -8T_cv_c+4-2nX=0.
    \label{eq:first_critical_X}
\end{equation}
Similarly, multiplying the second criticality condition by $v_c^4$
gives
\begin{equation}
    16T_cv_c-12+2n(2n+1)X=0.
    \label{eq:second_critical_X}
\end{equation}

Eliminating $T_cv_c$ between
Eqs.~\eqref{eq:first_critical_X} and
\eqref{eq:second_critical_X}, we obtain
\begin{equation}
    X
    =
    \frac{2}{n(2n-1)}.
    \label{eq:X_solution}
\end{equation}
Therefore,
\begin{equation}
    C_\chi v_c^{2-2n}
    =
    \frac{2}{n(2n-1)},
\end{equation}
or equivalently,
\begin{equation}
    v_c^{2n-2}
    =
    \frac{n(2n-1)}{2}C_\chi.
    \label{eq:vc_appendix}
\end{equation}
Using
\begin{equation}
    C_\chi
    =
    6\alpha4^{n-1}(1-\chi)^n,
\end{equation}
Eq.~\eqref{eq:vc_appendix} becomes
\begin{equation}
    v_c^{2n-2}
    =
    3\alpha n(2n-1)4^{n-1}(1-\chi)^n,
\end{equation}
which reproduces Eq.~\eqref{eq:vcPower}.

Substitution of Eq.~\eqref{eq:X_solution} into
Eq.~\eqref{eq:first_critical_X} gives
\begin{equation}
    T_cv_c
    =
    \frac{n-1}{2n-1}.
    \label{eq:Tcvc_appendix}
\end{equation}
Finally, evaluating Eq.~\eqref{eq:appendix_eos} at the critical
point yields
\begin{align}
    P_cv_c^2
    &=
    8T_cv_c
    -
    2
    +
    C_\chi v_c^{2-2n}
    \nonumber\\
    &=
    \frac{2(n-1)}{n},
    \label{eq:Pcvc_appendix}
\end{align}
which reproduces Eqs.~\eqref{eq:Tc} and \eqref{eq:Pc}.

\section{Alternative closure at fixed scale factor}
\label{app:fixed_a}

The criticality conditions depend on the prescription used to close
the nonflat state space. To illustrate this explicitly, consider
variations at fixed scale factor $a$, rather than at fixed $\chi$.
Define
\begin{equation}
    Y(v)
    \equiv
    \frac{4}{v^2}
    -
    \frac{k}{a^2}.
    \label{eq:Y_definition}
\end{equation}
The equation of state \eqref{eq:EOSa} then reads
\begin{equation}
    P(T,v;a,k)
    =
    \frac{8T}{v}
    -
    \frac{2}{v^2}
    +
    \frac{3\alpha}{2}Y^n.
    \label{eq:EOS_fixed_a}
\end{equation}
Since
\begin{equation}
    \left.
    \frac{\partial Y}{\partial v}
    \right|_a
    =
    -\frac{8}{v^3},
\end{equation}
the first derivative of the holographic contribution is
\begin{equation}
    \left.
    \frac{\partial}{\partial v}
    \left(
        \frac{3\alpha}{2}Y^n
    \right)
    \right|_a
    =
    -\frac{12\alpha n}{v^3}Y^{n-1}.
    \label{eq:first_derivative_fixed_a}
\end{equation}
Therefore,
\begin{equation}
    \left(
        \frac{\partial P}{\partial v}
    \right)_{T,a}
    =
    -\frac{8T}{v^2}
    +
    \frac{4}{v^3}
    -
    \frac{12\alpha n}{v^3}Y^{n-1}.
    \label{eq:dP_fixed_a}
\end{equation}

The second derivative is
\begin{align}
    \left(
        \frac{\partial^2P}{\partial v^2}
    \right)_{T,a}
    ={}&
    \frac{16T}{v^3}
    -
    \frac{12}{v^4}
    +
    \frac{36\alpha n}{v^4}Y^{n-1}
    \nonumber\\
    &+
    \frac{96\alpha n(n-1)}{v^6}Y^{n-2}.
    \label{eq:d2P_fixed_a}
\end{align}

Let $v_{c,a}$, $T_{c,a}$, and $Y_{c,a}$ denote the critical
quantities obtained under this alternative prescription. The first
criticality condition gives
\begin{equation}
    T_{c,a}
    =
    \frac{1}{2v_{c,a}}
    \left[
        1
        -
        3\alpha nY_{c,a}^{n-1}
    \right].
    \label{eq:Tc_fixed_a}
\end{equation}
Eliminating $T_{c,a}$ from the two criticality conditions yields
\begin{equation}
    1
    =
    3\alpha nY_{c,a}^{n-1}
    +
    \frac{24\alpha n(n-1)}{v_{c,a}^2}
    Y_{c,a}^{n-2},
    \label{eq:critical_fixed_a}
\end{equation}
where
\begin{equation}
    Y_{c,a}
    =
    \frac{4}{v_{c,a}^2}
    -
    \frac{k}{a^2}.
    \label{eq:Yc_fixed_a}
\end{equation}

Equation~\eqref{eq:critical_fixed_a} differs from the fixed-$\chi$
condition
\begin{equation}
    v_c^{2n-2}
    =
    \frac{n(2n-1)}{2}C_\chi.
\end{equation}
The two prescriptions coincide in the flat limit, $k=0$, but they
are generally inequivalent for $k\ne0$. This explicitly demonstrates
why the ensemble labels in the critical derivatives are essential.
The existence and stability of a critical point obtained from
Eq.~\eqref{eq:critical_fixed_a} require a separate analysis subject
to
\begin{equation}
    Y_{c,a}=H_c^2>0,
    \qquad
    T_{c,a}>0.
\end{equation}

\bibliographystyle{apsrev4-1}
\bibliography{biblio.bib}

\end{document}